\documentclass[preprint,12pt]{elsarticle}
\usepackage{amssymb}
\usepackage{graphicx}
\usepackage{amsmath,amsfonts}

\begin{document}

\begin{frontmatter}

\title{Operational Optimal Ship Routing Using a Hybrid Parallel Genetic Algorithm}

\author[UoP]{O. T. Kosmas}
\ead{odykosm@uop.gr}

\author[UoP]{D. S. Vlachos}
\ead{dvlachos@uop.gr}

\address[UoP]{Laboratory of Computer Sciences,\\
Department of Computer Science and Technology,\\
Faculty of Sciences and Technology, University of Peloponnese\\
GR-22 100 Tripolis, Terma Karaiskaki, GREECE}

\begin{abstract}
Optimization of ship routing depends on several parameters, like
ship and cargo characteristics, environmental factors, topography,
international navigation rules, crew comfort etc. The complex nature
of the problem leads to oversimplifications in analytical
techniques, while stochastic methods like simulated annealing can be
both time consuming and sensitive to local minima. In this work, a
hybrid parallel genetic algorithm - estimation of distribution
algorithm is developed in the island model, to operationally
calculate the optimal ship routing. The technique, which is
applicable not only to clusters but to grids as well, is very fast
and has been applied to very difficult environments, like the Greek
seas with thousands of islands and extreme micro-climate conditions.
\end{abstract}


\end{frontmatter}

\section{Introduction}
\label{intro}
Optimization of ship routing is closely related to
both ship characteristics and environmental factors and has a
significant influence on economical, safety and comfort
considerations. Ship size, speed capability and type and scheduling
of cargo are important considerations in the route selection process
prior to sailing and the surveillance procedure while underway.
Ship's characteristics identify its vulnerability to adverse
conditions and its ability to avoid them while cargo type and
scheduling identify the safety standards that have to be fulfilled
and the economical impact of a certain route \cite{vlachos_EOS_69_649_03}.
\par On the other hand, environmental factors of importance to ship routing are
those elements of the atmosphere and ocean that may produce a change
in the status of a ship transit. In ship routing, consideration is
given to wind, waves, fog and ocean currents. While all of the
environmental factors are important for route selection and
surveillance, optimum routing is normally considered attained if the
effects of wind and waves can be optimized. More details about the
effect of environmental factors can be found in \cite{vlachos_ANACM_3_547_04}.
\par The problem of calculating an optimal or near optimal ship route
grows very fast in complexity with the inclusion of several
realistic constraints, like the existence of islands, international
or national navigation rules, micro-climate and local parameters,
etc. Any trial for an efficient analytical solution soon will be
locked in oversimplifications, while heuristics have been proved,
during the last years, capable to achieve acceptable solutions in
such complicated problems. See for example the Traveling Salesman
Problem (\cite{chatterjee_EJOR_93_490_96} and \cite{katayama_MCM_31_197_00}), the
Time-Table Problem (\cite{burk_IEEETEC_3_64_99}), the Quadratic Assignment
Problem (\cite{ahuja_COR_27_917_00},\cite{nissen_IEEETNN_5_66_94},\cite{schnecke_IEEETEC_1_266_97}
and \cite{tate_COR_22_73_95}), the Job Shop Scheduling Problem
(\cite{cheng_CIE_30_983_96},\cite{cheng_CIE_36_343_99},\cite{della__COR_22_15_95},\cite{wang_COR_28_585_01}
and \cite{zhou_CIE_40_191_01}), the Airline Scheduling Problem
(\cite{ozdemir_IS_133_165_01}) and others. The best results found for many
practical or academic optimization problems are obtained by hybrid
algorithms. Combination of algorithms such as descent local search
\cite{papadimitriou_Book_COAC_PH_82}, simulated annealing
\cite{kirkpatrick_S_220_4598671_83}, tabu search \cite{glover_Book_TS_KAP_97} and
evolutionary algorithms have provided very powerful search
algorithms.
\par Beyond the complexity of the optimal ship routing problem, one has
to deal with the necessity to build a system that can respond to
user requests in a reasonable amount of time. A user may identify
himself which means that the system has registered information about
the user's ship. After that, the user gives its departure and
arrival locations with the desired departure time. The system
responds with the optimal route calculated, taking into account the
user's requirements (for example, small and medium ships are more
interested in safety and comfort, while storeships are interested in
fuel consumption and cargo scheduling). In any case, the response
should be provided immediately (after a few seconds) otherwise, it
is very likely that users exhibit disinclination in using the
system.
\par Finally, one has to take special care for the well-known
exploration-exploitation trade-off. Exploration is needed to ensure
every part of the search space is searched thoroughly in order to
provide a reliable estimate of the global optimum. Exploitation is
important since the refinement of the current solution will often
produce a better solution. Consider the case where in a given
landscape with mountains and valleys, one wants to calculate the
shortest path between two given points. In order to roughly locate
this path, one has to observe the landscape from a long distance, so
he can obtain a general idea of all the possible routes in the
landscape. Thus, in our example, exploration means to observe from
far away. On the other hand, after locating the possible optimal
route, on has to take a closer look at the landscape in order to
locate local obstacles and obtain a fine tuning of his path. Thus,
exploitation in our example means to observe nearby. It is clear
that exploration and exploitation are two competing tasks.
Population-based heuristics (where genetic algorithms
\cite{holland_Book_ANAS_UMP_75} and estimation of distribution algorithms
\cite{larranaga_Book_EDANTEC_KAP_01} are found) are powerful in the exploration of
the search space, and weak in the exploitation of the solutions
found.
\par The integrate solution of the aforementioned difficulties
(lots of constraints, immediate response and exploration-exploitation
trade off) in the optimal ship routing problem is the aim of this work.
The material is organized as follows:
In section \ref{osrp}, a brief review of the optimal ship routing
problem is given. In section \ref{na}, a detailed description of the
proposed algorithm is given. Section \ref{er} summarizes experimental
results, while the conclusions of the implementation of the new
algorithm are presented in section \ref{conc}.
\section{The optimal ship routing problem}
\label{osrp}
\begin{figure}[!t]
\centering
\includegraphics[width=\textwidth]{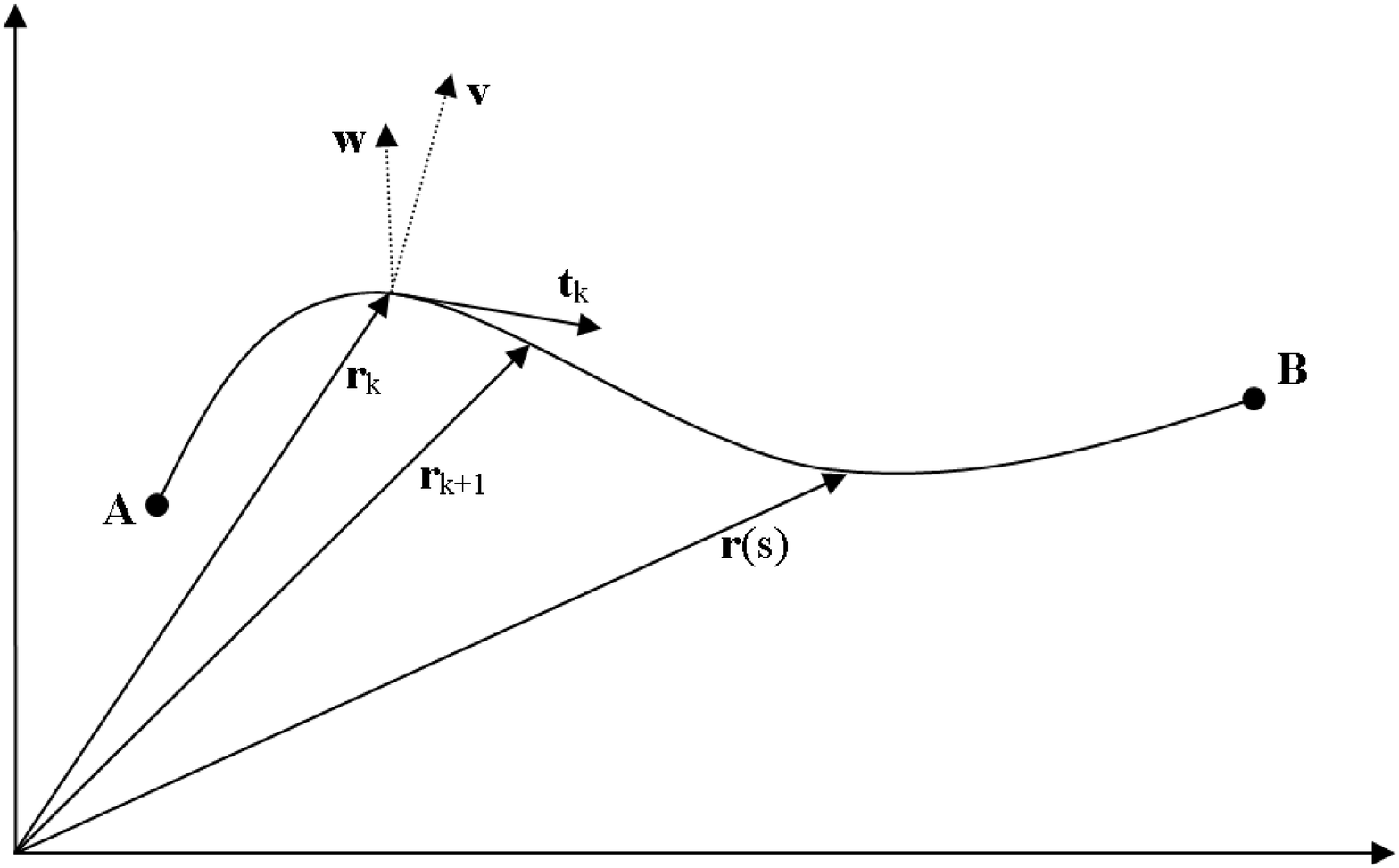}
\caption{A route from point A to point B. $r(s)$ is the
parametrization of the route, and $w,v$ are the wave and wind
vectors respectively and $t_k$ is the tangent to the route at point
$k$.} \label{fig-form}
\end{figure}

Let us assume that an initial route of a ship is represented by a
smooth curve $\vec{r}(s)$  (Fig. \ref{fig-form}) where the parameter
$s$ is the arc length measured from some fixed point A (initial
point of the ship route). Then, the tangent vector of the curve of
the ship's route in the point of question is defined as
\begin{equation}
\vec{t}=\frac{\dot{\vec{r}}}{|\dot{\vec{r}}|}=\frac{d\vec{r}}{ds}
\end{equation}
where  $\dot{\vec{r}}$ is the ship's velocity.

We also assume that the moving ship is subject to the influence of
the wave height and direction represented by the vector $\vec{w}$
and the wind speed and direction represented by the vector
$\vec{v}$.

Under the above assumptions we define a route cost (a scalar
quantity) assigned at every possible route between two points. The
route cost includes a weighted combination of the voyage time and
the safety (or comfort) of the voyage. The total cost $S$ is given
by
\begin{equation}
S=aT+(1-\alpha )C\label{equ_cost}
\end{equation}
where $T$ is the total voyage time and $C$ is a scalar
characterizing the safety (comfort) of the voyage. The weight
$\alpha$ can be tuned by the user depending on his demands. Note
here that when $\alpha = 1$, then the only optimization parameter is
the voyage time while when $\alpha = 0$, the only optimization
parameter is the crew comfort. The scalar $C$ is calculated as a
line integral over the route by the following way (up to the linear
approximation):
\begin{equation}
C=\int_{A}^{B}{(\vec{v}^{T}Z_{v}+\vec{w}^{T}Z_{w})\vec{t}ds}
\end{equation}
where $\vec{v}$ is the wind vector, $\vec{w}$ is the wave height
vector and $Z_{v}$ and $Z_{w}$ are tensors which characterize the
ship response to wind and wave, respectively. The calculation of the
total voyage time $T$, is a bit more complicated, since both wind
and waves can alter the speed of the ship. In general we can write
that
\begin{equation}
|\dot{\vec{r(r)}}|=u+F(\vec{v},\vec{w},\vec{t})
\end{equation}
where $u$ is the speed of the ship in zero wind and $F$ is a
function that depends on ship characteristics, wind, wave and
direction of the ship movement. For simplicity in the present work,
we assume that $F = 0$. Moreover, we assume that a candidate route
can be represented as a set of way-points, while the path between
two successive way-points is always a straight line (or a great
circle on the globe). Obviously, the coordinates of the way-points
is the objective of the search process. Without loss of generality,
we assume that the departure and arrival points lay on the
horizontal axis (if not, we can always rotate the coordinate
system). In order to minimize the search space, we can assume that
the horizontal positions of the way-points are fixed, and the
objective of the search process is the determination of the vertical
coordinates of the way-points. This simplification not only is
acceptable but is indicated from the fact that the environmental
parameters who affect the optimal route are known in a grid (the
grid of the forecasting model used to account for the state of the
sea during the next hours or days). Therefore, more than one
way-points in the neighborhood of a grid point cannot be optimized
efficiently (since no extra information is available for these
points from the forecasting model).
\section{The new algorithm}
\label{na}
As it was mentioned before, there are three main difficulties that
one has to overcome, in order to produce an efficient and
operational algorithm for the optimal ship routing. In the
following, a detailed view of how the proposed algorithm handles
them is presented.
\subsection{Constraints in ship routing}
\label{csr}
Constraints in ship routing arise from the existence of obstacles
(islands) and general international or national navigation rules.
The penalty encoding method perhaps is the most popular approach
used in Genetic Algorithms for constrained optimization problems
because of its simplicity and ease of implementation
(\cite{chau_AC_13_481_04},\cite{contreras_LNCS_3533_547_05},\cite{deb_CMAME_186_311_00},\cite{gen_Proc_IEEECEC_804_96},\cite{ji_JSJU_40_811_05}
and \cite{michalewicz_CIE_30_851_96}). On the other hand, one can take extra
actions in order to limit the members of the population in the
feasible region of the search space. There are several techniques
that have been proposed, from modified mutation and crossover
operation, with the property that they only produce feasible
offsprings from feasible parents to the death penalty method, in
which a member is destroyed, if it violates a certain constrain. But
in the case of the optimal ship routing this method is almost
unapplicable. The reason is the large numbers of constraints
(islands) which makes extremely difficult to design mutation and
crossover operators which produce feasible offsprings. A nice short
review of proposed methods for handling constraints in genetic
algorithms with selected references can be found in
\cite{farmani_IEEETEC_7_445_03}.

\subsubsection{General penalty formalism}
\label{gpf}
\begin{figure}[!t]
\centering
\includegraphics[width=\textwidth]{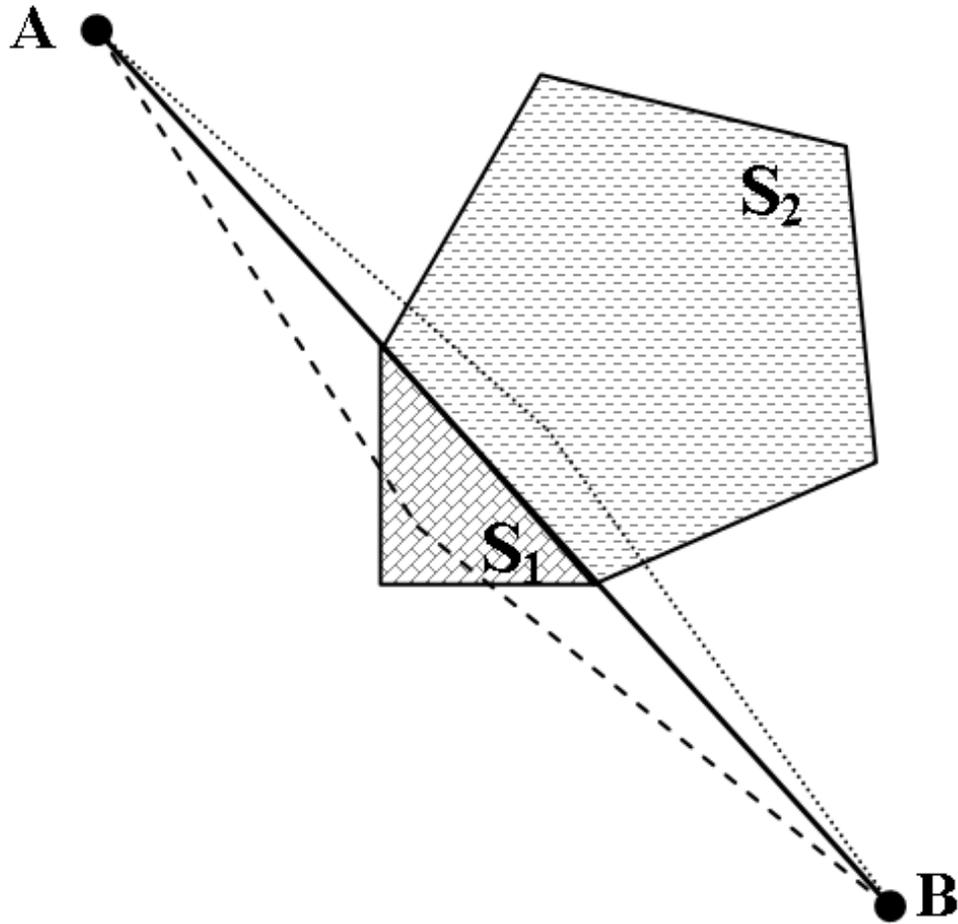}
\caption{A route from point A to point B which crosses an obstacle
(solid route). The obstacle is divided by the route into two areas,
$S_1$ and $S2$. The ratio of the smaller area to the larger one, is
decreased in the right direction (dashed route) while is increased
in the opposite direction (dotted route).} \label{fig_pen}
\end{figure}

In Fig. \ref{fig_pen} a route between two points $A$ and $B$ is
shown. This route crosses an obstacle and divides it into two parts,
$S_1$ and $S_2$. In order to assign a constrain with the obstacle,
we calculate the ratio
\begin{equation}
h=-\frac{min\{S_1,S_2\}}{max\{S_1,S_2\}}=-\frac{S_1}{S_2}
\end{equation}
If the route does not cross the obstacle, then the area of $S_1$ is
zero and that of $S_2$ is the area of the whole obstacle. Thus, the
parameter $h$ takes the value $0$ if the route does not cross the
obstacle and a negative value otherwise. In the case where the
obstacle is divided in exactly two equal areas, $h$ takes the value
$-1$, which is the smaller value that can be assigned to $h$. It is
clear now, that whatever are the actions of the optimization
algorithm, the value of $h$ has to be increased. Moreover, this
encoding of the constraint shows us the direction of the change that
have to be induced in the route. As it is shown in Fig.
\ref{fig_pen}, the dashed route decreases the value of $h$, while
the dotted one increases it.
\par Consider now that we have $N$ obstacles and for each one of them we calculate the parameter $h$.
Then, for a feasible solution the following equation must hold:
\begin{equation}
h^i = 0 \;,\; i=1,...,N\label{equ_eqcon}
\end{equation}
On the other hand, there is another type of constraints which are
caused by a practical inability of a ship to follow abrupt changes
in the movement direction. If we force the ship to take sharp turns,
then this might cause safety problems especially in heavy seas.
Consider a route which is composed by straight lines joining the way
points $P_0,P_1,...,P_M$, where $P_0$ and $P_M$ is the starting and
ending point respectively. For each part of the route, we calculate
the direction vector $d^i$
\begin{equation}
d^i=\frac{\mathbf{P}^i-\mathbf{P}^{i-1}}{|\mathbf{P}^i-\mathbf{P}^{i-1}|}\;,\;i=1,...,M
\end{equation}
where $\mathbf{P}^i$ is the position vector of point $P_i$. Then,
for a feasible solution we want that
\begin{equation}
g^k=cos^{-1} \left( d^k \cdot d^{k+1}\right) -\phi _{max} \geq 0
\;\;,\;k=1,...,M-1 \label{equ_incon}
\end{equation}
where $\phi _{max}$ is the maximum allowed turn that the ship can
take. Thus, for a given route $\vec{x}$ we have two types of
constraints, i.e. $N$ equalities $h^i (\vec{x})$ and $M-1$
inequalities $g^k (\vec{x})$ (equations (\ref{equ_eqcon}) and
(\ref{equ_incon}) respectively). Note here that a route is
represented by a vector containing the $M+1$ way points
$P_0,P_1,...,P_M$. If now $\delta(x)$ is the Dirac delta function
and $u(x)$ is the step function with
\begin{equation}
u(x)=\Big\{ \begin{array}{lll}1 & , & x\geq 0 \\ 0 & , & x<0
\end{array}
\end{equation}
then the ideal penalty function for a configuration $\vec{x}$ is
given by:
\begin{equation}
p(x)=\frac{1-u(g(\vec{x}))}{u(g(\vec{x}))}+\frac{1}{\delta(h(\vec{x}))}
\end{equation}
where:
\begin{itemize}
\item the first term is zero, if the inequalities $g$ hold and tends
to infinity otherwise
\item the second term is zero, if equalities $h$ hold and tends to
infinity otherwise
\end{itemize}
\subsubsection{Smooth penalty formalism}
\label{spf}
\begin{figure}[!t]
\centering
\includegraphics[width=\textwidth]{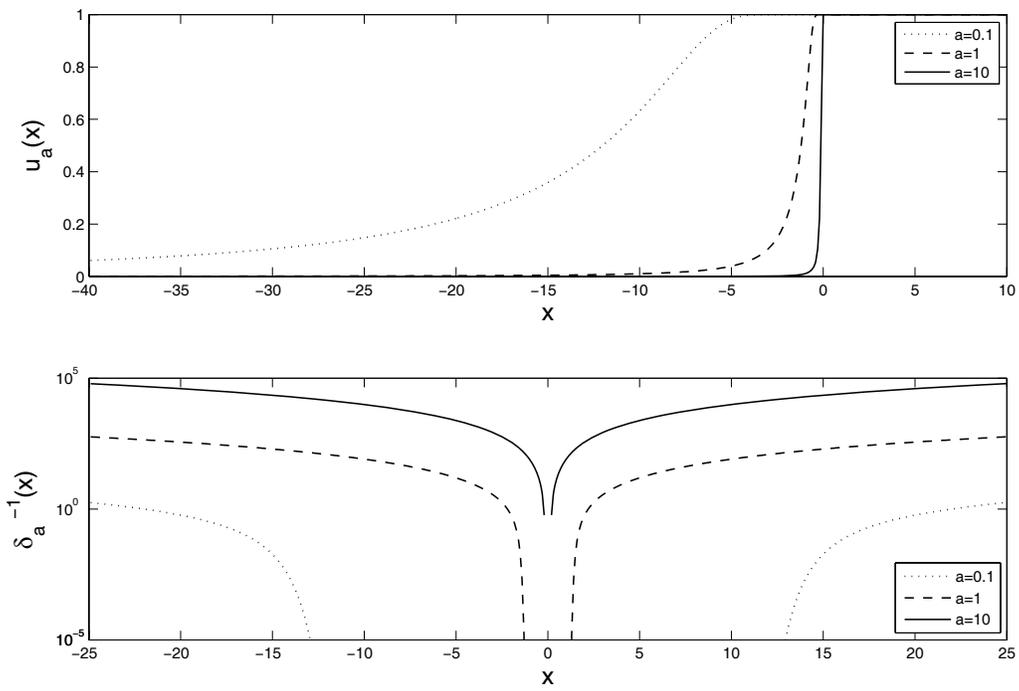}
\caption{The functions $u_a(x)$ and $\frac{1}{\delta _a(x)}$ for
different values of the parameter $a$.} \label{fig_ud}
\end{figure}

Instead of using the discontinuous functions $u(x)$ and $\delta
(x)$, we can approach them with the $C^{\infty}$ functions
\begin{equation}
\hat{u}_a(x)=\Big\{\begin{array}{lll}1-e^{-1\frac{1}{(ax)^2}}&,&x< 0
\\ 1&,x\geq 0\end{array}\label{equ_u}
\end{equation}
and
\begin{equation}
\frac{1}{\hat{\delta}_a (x)} =\Big\{
\begin{array}{lll} \frac{1}{e^{\frac{1}{(ax+1)^2}}-1} &,&x<-\frac{1}{a} \\
0&,&-\frac{1}{a}\leq x\leq \frac{1}{a}
\\\frac{1}{e^{\frac{1}{(ax-1)^2}}-1} &,&x>\frac{1}{a}\label{equ_d}
\end{array}
\end{equation}
where both $\hat{u}_a(x)$ and $\hat{\delta}_a(x)$ tend to $u(x)$ and
$\delta (x)$ respectively when $a\rightarrow \infty$. The penalty
function is given now
\begin{equation}
P(\vec{x})=\frac{1-\hat{u}_a(g(\vec{x}))}{\hat{u}_a(g(\vec{x}))}+\frac{1}{\hat{\delta}_b(h(\vec{x}))}
\end{equation}
for some given $a,b$. It can be easily shown that both functions
$\hat{u}_a$ and $\hat{\delta}_b ^{-1}$ are $C^{\infty}$ functions
everywhere in $R$. Furthermore, an obvious advantage of these two
functions is that although they are smooth, they add no penalty at
all to feasible solutions. Figure \ref{fig_ud} shows the functions
$\hat{u}_a$ and $\hat{\delta}_a^{-1}$ for different values of $a$.
\subsubsection{Complex cost function}
\label{ccf}
Let us assume now that the problem under consideration can be
reduced to the minimization of the everywhere positive function
$S(\vec{x})$ given by equation (\ref{equ_cost}). Consider the
complex function $\Lambda(\vec{x})$:
\begin{equation}
\Lambda (\vec{x})=S(\vec{x})+i\cdot P(\vec{x})
\end{equation}
Minimization of $|\Lambda(\vec{x})|$ is now equivalent to our
problem. A feasible solution to our problem must lie on the real
axis (in our case it is the x-configuration space). This can be
smoothly achieved by considering the generalized cost function $E$
given by:
\begin{equation}
E(\vec{x})=|\Lambda(\vec{x})|\cdot \rho(\lambda , \Lambda(\vec{x})
)\label{equ_modcost}
\end{equation}
where the multiplicative term $\rho$ is given
\begin{equation}
\rho (\lambda , \Lambda )=\Big\{ \begin{array}{lll}
1+\frac{1}{e^{\frac{1}{(\lambda Im(\Lambda)-1)^2}}-1} &,&Im(\Lambda
)>\frac{1}{\lambda} \\1 &,& 0\leq Im(\Lambda ) \leq
\frac{1}{\lambda}
\end{array}
\end{equation}
It is clear now that we restrict the feasible space in the zone
$0\leq P(\vec{x}) \leq \frac{1}{\lambda}$. Finally, a sort of
annealing is introduced here, pushing $\lambda$ to $\infty$ and thus
moving the solution to the real axis which is the feasible space of
the problem.

\subsection{Operational principles}
\label{op}
\begin{figure}[!t]
\centering
\includegraphics[width=\textwidth]{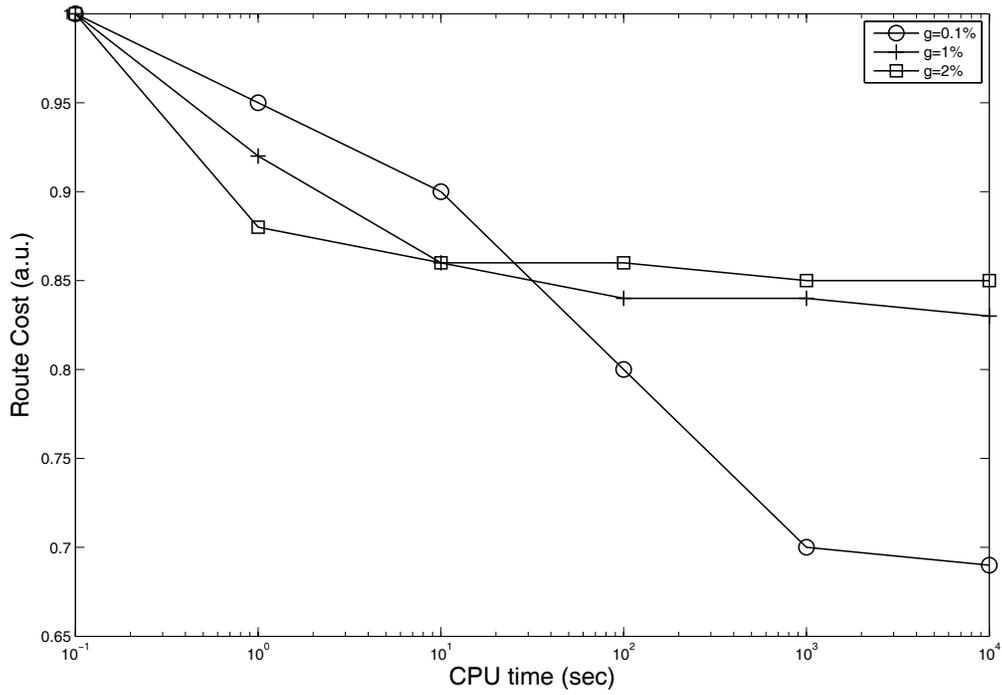}
\caption{The application of a typical genetic algorithm with 8-bit
encoding for each way point, $10$ way points for each route and the
penalty method described in section \ref{ccf}. In these experiments, the
parameter $\lambda$ of equation (\ref{equ_modcost}) had the same
value as the parameter $a$ of equations (\ref{equ_u}) and
(\ref{equ_d}). Finally, the experiment was carried on a
$Pentium^{\circledR}$ $4$ at $2.4GHz$. The parameter $g$ is the percentage
of the increase of $\lambda$ in every generation.} \label{fig_res1}
\end{figure}

Having in mind that a route may contains $10$ to $20$ way points as
it will be explained later and the search space is the Eastern
Mediterranean with hundreds of islands, we find that a candidate
solution has to fulfill hundreds of constraints, which makes the
implementation of the algorithm to a single computer impractical.
Fig. \ref{fig_res1} shows the application of a typical genetic
algorithm with 8-bit encoding for each way point, $10$ way points
for each route and the penalty method described earlier. In these
experiments, the parameter $\lambda$ of equation (\ref{equ_modcost})
had the same value as the parameter $a$ of equations (\ref{equ_u})
and (\ref{equ_d}). Finally, the experiment was carried on a
$Pentium^{\circledR}$ $4$ at $2.4GHz$. This experiment tests the quality
of the solution depending on the annealing rate and thus on the cpu
time. It is clear that the algorithm is sensitive to the rate of
annealing, while for slow annealing, although we obtain good
solutions, the computational time is inhibitory.
\par
\begin{figure}[!t]
\centering
\includegraphics[width=\textwidth]{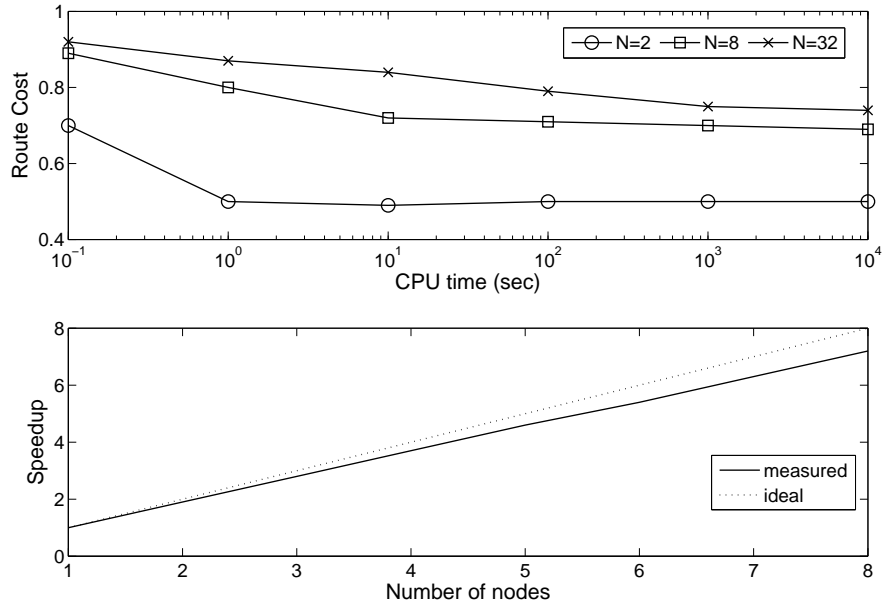}
\caption{Precalculation of obstacle bypasses facilitate the
convergence when the number of obstacles $N$ is small, but very
soon, with increasing $N$ the cpu time becomes inhibitory. The
speedup of this implementation, although close to ideal, cannot
solve the problem.} \label{fig_res2}
\end{figure}

 The use of a parallel system or a grid of computers is thus
necessary in order to obtain practical response times. Two
approaches have been tested. The first one is to facilitate the
searching mechanism by pre-calculating all the possible bypasses of
the obstacles between the first and last way points. For each one of
the pre-calculated routes, we generate a population of routes that
are close to the initial ones. Each population is evolved using a
death penalty mechanism for handling the constraints. Finally, the
best solution is selected. The results of this approach are shown in
Fig. \ref{fig_res2}a. The cost of the final solution is drawn as a
function of time, for several numbers of obstacles between the first
and last way point. In Fig. \ref{fig_res2}b the speedup of the
implementation of the method in a cluster with $8$ nodes is shown.
Since for every obstacle added between the first and final way
point the number of possible bypasses are doubled, this method
fails for long routes which have to bypass tenths of islands.
\par The second approach to the problem is based on the synergetic
action of two algorithms, the genetic algorithm and the estimation
of distribution algorithm, as it will be explained in details in the
following paragraphs.

\subsection{The exploration exploitation trade-off}
\label{eeto}
The first difficulty which arises in the optimal ship routing is the
large dimension of the search space. Consider the case where we want
to approximate the optimal solution between two points which lie on
the $x$-axis with a set of $M$ way points between the first and the
last one. As it was mentioned before, it is reasonable to fix the
$x$-coordinate of the way points and try to optimize the
$y$-coordinates of the way points. Thus the search space is
$M$-dimensional. Since we are using forecasting data for the sea
state and wind and a typical grid size for such forecasting models
is $0.1^o$, a set of $20$ way points will be adequate to cover every
small or medium size ship voyage. Moreover, it is reasonable to
bound the $y$-coordinates of the way points in a rectangle with the
size of the edge parallel to the $x$-axis to be the distance between
the first and last way point and the size of the edge parallel to
the $y$-axis to be double. If we use a binary encoding for every
$y$-coordinate with $n$-bits, then the search space is divided in
$M\cdot 2^n$ cells with surface area $\Delta S$ given
\begin{equation}
\Delta S=\frac{d^2}{M\cdot 2^{n-1}}
\end{equation}
where $d$ is the distance between the first and last way points. The
value of $\Delta S$ is a measure of the exploration-exploitation of
the genetic algorithm. A large value for $\Delta S$ will soon
produce a solution which locates in general the limits of the
$y$-coordinates, while a small value for $\Delta S$ will produce a
fine tuning of a given route.
\par This observation lead us to the construction of several
population with different size in the bit encoding of the
$y$-coordinates. Moreover, population are organized in a
hierarchical network, in which members of populations with $l$ bits
per coordinate in the bit encoding can migrate only to populations
with $l'\geq l$ bits per coordinate. The result of this approach is
that the populations in the upper levels of the network converge fast to
solutions (due to the small number of cells in the search space that
they have to look in) and this information is forwarded in
population of lower levels in the network, where fine tuning
(exploitation) is performed.

\subsection{Hybrid GA-EDA Algorithms}
\label{hgaeda}
The final part of the proposed algorithm is the way that members
from one population can migrate to another. The hybrid GA-EDA
(genetic algorithm - estimation of distribution algorithm) technique
is used. The benefits of this approach are (a) the saving of
communication time between parallel processes and (b) the inclusion
of an extra searching mechanism. More specifically, since members
from one population have to migrate in order to carry information
about the search space, the amount of data that are transferred
between processes affect dramatically the speedup of the parallel
algorithm (the communication time is at least four orders of
magnitude bigger than the processing time). Thus it is by far more
efficient to migrate the distribution function of the genes of the
members than the members themselves.
\par On the other hand, the original objective is
to get benefits from both approaches. The main difference from these
two evolutionary strategies is how new individuals (offsprings) are generated.
Our new approach generates two groups of offspring
individuals, one generated by the GA mechanism and the other by EDA
one. GAs use crossover and mutation operators as a
mechanism to create new individuals from the best individuals of the
previous generation. On the other hand, EDA builds a probabilistic
model with the best individuals and then samples the model to
generate new ones. Population $p+1$ is composed by the best overall
individuals from (i) the past population, (ii) the
GA-evolved offspring, and (iii) EDA-evolved offspring. The
individuals are selected based on their fitness function. This
evolutionary schema is quite similar to Steady State GA in which
individuals from one population, with better fitness than new
individual from the offspring, survive in the next one.

\section{Experimental results}
\label{er}
\begin{figure}[!t]
\centering
\includegraphics[width=\textwidth]{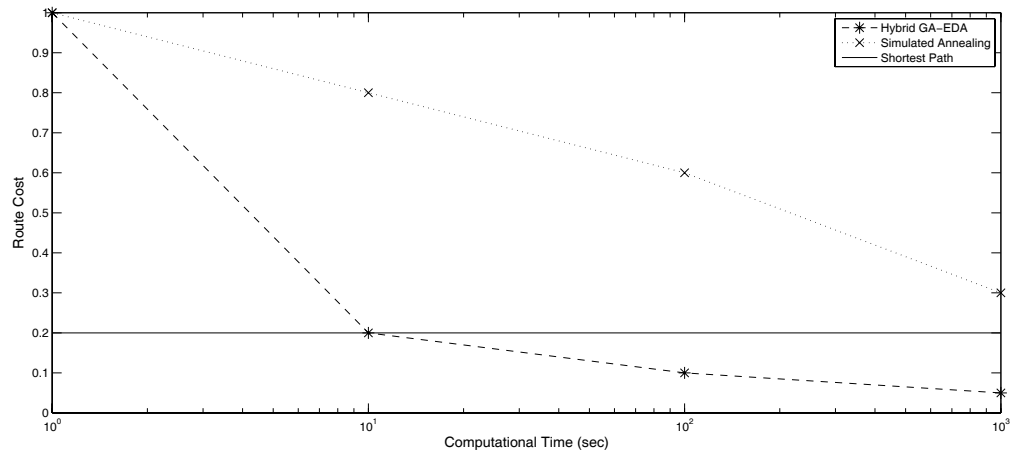}
\caption{The cost of the optimal route from the port of Thessaloniki $(40.5197N,$ $22.9709E)$ to the port of Ag. Nikolaos ($35.1508N,\;25.7227E$), calculated by the proposed GA-EDA algorithm ($*$) compared to the cost of the route calculated using simulated annealing ($\times$). The cost of the shortest path is drawn too (solid line).} \label{fig_res3}
\end{figure}

In order to test the proposed algorithm, the optimal route is calculated from the
port of Thessaloniki ($40.5197N,\;22.9709E$) to the port of Ag.
Nikolaos ($35.1508N,\;25.7227E$). Forecast data are taken from
climate databases. There are more than 20 islands which give more than $2^{20}$ ways to bypass them.
Fig. \ref{fig_res3} shows the cost of the calculated route as a function of computational time and this is compared to the cost of a route calculated using simulated annealing. The cost of the shortest path is also drawn in the same figure. The proposed algorithm gives operational results in at least 2 orders of magnitude less time than simulated annealing. Finally these results have been reproduced several times using variable environmental conditions.

\section{Conclusion}
\label{conc}
In this work a hybrid evolutionary method based on genetic
algorithms and estimation of probability distribution algorithm has
been designed and implemented to deal with the problem of optimal
ship routing. The large number of constraints lead us to the
development of a parallel system in order to produce good solution
in reasonable times and thus integrate this method in an operational
advisory system. The basic parts of the method are:
\begin{enumerate}
\item A smooth penalty function has been constructed to deal with
constraints which by using an annealing mechanism forces the
searching to be bounded in the feasible area of the search space.
\item In order to account with the dependence of the convergence of
the algorithm on the annealing rate, different populations evolve
with different annealing rates. This, in combination with the
exchange of groups of members between the populations, unlocks the
algorithm from the local minima.
\item In order to handle the exploration exploitation trade-off,
different populations evolve with different binary encoding, while
all of them cover the whole searching space. Populations with smaller
size in bit representation of the solution have a better exploration
performance, while those with larger size have better exploitation
capabilities.
\item Information between populations are exchanged using the
estimation of distribution algorithm. This, not only decreases the
communication cost between the nodes of the parallel system, but
includes an extra mutation operation which improves the searching
capabilities of the algorithm.
\end{enumerate}
Experimental results both from simulation and real data show that
the system meets its specifications and thus could be used in
operational mode.

\section*{Acknowledgment}
This paper is part of the 03ED51 research project, implemented
within the framework of the "\emph{Reinforcement Programme of Human
Research Manpower}" (\textbf{PENED}) and co-financed by National and
Community Funds (25\% from the Greek Ministry of Development-General
Secretariat of Research and Technology and 75\% from E.U.-European
Social Fund).

\bibliographystyle{elsarticle-num}
\bibliography{par_EDA}

\end{document}